\documentclass[12pt,reqno]{amsart}
\usepackage{amsthm,amsfonts,amssymb,euscript}

\newcommand{\bea}{\begin{eqnarray}}
\newcommand{\eea}{\end{eqnarray}}
\def\beaa{\begin{eqnarray*}}
\def\eeaa{\end{eqnarray*}}

\def\c{\cdot}
\def\id{\mbox{id}}

\newcommand{\nn}{\nonumber}

\def\ga{\gamma}

\def\de{\delta}
\def\De{\Delta}

\def\la{\lambda}

\def\si{\sigma}

\def\om{\omega}

\def\nab{\nabla}

\def\II{{\mathcal I}}

\def\x{{\bf x}}
\def\xib{{\underline \xi}}
\def\tb{{\underline t}}

\def\SSS{{\Bbb S}}
\def\RRR{{\Bbb R}}

\begin{document}
\theoremstyle{plain}
  \newtheorem{theorem}[subsection]{Theorem}
  \newtheorem{conjecture}[subsection]{Conjecture}
  \newtheorem{proposition}[subsection]{Proposition}
  \newtheorem{lemma}[subsection]{Lemma}
  \newtheorem{corollary}[subsection]{Corollary}

\theoremstyle{remark}
  \newtheorem{remark}[subsection]{Remark}
  \newtheorem{remarks}[subsection]{Remarks}

\theoremstyle{definition}
  \newtheorem{definition}[subsection]{Definition}

\newcommand{\Lp}[2]{\left\Vert \, #1 \, \right\Vert_{#2}}
\newcommand{\lp}[2]{\Vert \, #1 \, \Vert_{#2}}

\newcommand{\spc}{\hspace{.3cm}}
\newcommand{\ret}{\vspace{.1cm}}
\title
{On the uniqueness of solutions to the Gross-Pitaevskii hierarchy}
\author{Sergiu Klainerman}
\address{}
\email{}

\author{Matei Machedon}
\address{}
\email{}
\thanks{}
\date{}
\dedicatory{}
\commby{}
\begin{abstract}
The purpose of this note is to give a new proof of uniqueness of the Gross- Pitaevskii
hierarchy, first established in \cite{ESY1},
  in a different space,   based on space-time
  estimates  similar  in spirit to those
of \cite{KM}. 

\end{abstract}
\maketitle

\section{Introduction}

The Gross-Pitaevskii hierarchy refers to a sequence of functions
$\gamma^{(k)}(t, \bold x_k, \bold x'_k)$, $k = 1, 2, \cdots$, 
where $t \in \RRR$, $\bold x_k=(x_1, x_2, \cdots, x_k)
\in \RRR^{3k}$, $\bold x'_k=(x'_1, x'_2, \cdots, x'_k)
\in \RRR^{3k}$ which are symmetric,  in the sense that
\[
\gamma^{(k)}(t, \bold x_k, \bold x'_k)
= \overline{\gamma^{(k)}(t, \bold x'_k, \bold x_k)}
\]
and
\begin{equation}\label{symm}
\gamma^{(k)}(t, x_{\sigma (1)}, \cdots x_{\sigma (k)}, x'_{\sigma (1)}, \cdots x'_{\sigma (k)} )
=\gamma^{(k)}(t, x_1, \cdots x_k, 
x'_1, \cdots x'_k)
\end{equation}
for any permutation $\sigma$,
and satisfy the Gross-Pitaevskii infinite linear hierarchy of equations, 
\begin{equation}\label{G-P}
\left( i \partial_t + \Delta_{\bold x_k} - \Delta_{\bold x'_k}\right)\gamma^{(k)}\\
=\sum_{j=1}^{k}B_{j, \, k+1}(\gamma^{(k+1)}).
\end{equation}
with prescribed initial conditions 
\beaa
\gamma^{(k)}(0, \bold x_k, \bold x'_k)
=\gamma_0^{(k)}(\bold x_k, \bold x'_k).
\eeaa
Here  $\Delta_{\x_k}$,  $\Delta_{\x_k'}$  refer to the standard 
Laplace operators with respect to the variables
$\x_k, \x_k'\in \RRR^{3k}$ and  the operators  $B_{j, \,k+1}= B^1_{j, \,k+1} - B^2_{j, \,k+1}$ are defined according to,
\beaa
&& B^1_{j, \, k+1}(\gamma^{(k+1)})(t, \bold x_k, \bold x'_k) \label{Bs}\\
&& =\int\int \delta(x_j - x_{k+1}) \delta (x_j - x'_{k+1})\gamma^{(k+1)}(t, \bold x_{k+1}, \bold x'_{k+1})\, d x_{k+1}\, d x'_{k+1} \\
&& B^2_{j, \, k+1}(\gamma^{(k+1)})(t, \bold x_k, \bold x'_k)\notag\\
&& =\int\int \delta(x'_j - x_{k+1}) \delta (x'_j - x'_{k+1})\gamma^{(k+1)}(t, \bold x_{k+1}, \bold x'_{k+1})\, d x_{k+1}\, d x'_{k+1}.
\eeaa
In other words $B^1_{j, \, k+1}$, resp. $B^2_{j, \, k+1}$,  acts on
 $\gamma^{(k+1)}(t, \bold x_{k+1}, \bold x'_{k+1})$  replacing 
both variables  $x_{k+1}$ and $x'_{k+1}$    by $x_j$,   resp $x_j'$. 
We shall also make use of the operators,
\beaa
B^{k+1}=\sum_{1\le j \le k}B_{j, \, k+1}
\eeaa

One can easily verify that a particular  solution to \eqref{G-P} is given by,
\bea
\gamma^{(k)}(t, \bold x_k, \bold x'_k)= \prod_{j=1}^{k} \phi(t, x_j) \overline \phi(t, x'_j)
\label{eq:special-sol}
\eea
where each $\phi$ satisfies the non-linear Schr\"odinger equation 
in 3+1 dimensions
\begin{equation}\label{eq:NLS}
 \left(i \partial_t + \Delta \right)\phi = \phi |\phi|^2 ,\quad \phi(0, x)=\phi(x)
\end{equation}

In  \cite{ESY1} 
L. Erd\"os,  B. Schlein and H-T Yau provide a rigorous derivation
 of the cubic non-linear Schr\"odinger equation \eqref{eq:NLS}
 from the quantum dynamics
of many body systems.
 An important  step in their program is  to  prove 
uniqueness to solutions of \eqref{G-P}  corresponding to the special
 initial conditions
\bea
\gamma^{(k)}(0, \bold x_k, \bold x'_k)=\gamma_0^{(k)}(\bold x_k, \bold x'_k)=\prod_{j=1}^{k} \phi( x_j) \overline \phi( x'_j)\label{special-IC}
\eea
with $\phi \in H^1(\Bbb R^3)$.
To state precisely  the uniqueness result
 of \cite{ESY1},  denote $S_j=(1-\Delta_{x_j})^{1/2}$,
$S'_j=(1-\Delta_{x'_j})^{1/2}$ 
and $S^{(k)}= \prod_{j=1}^k S_j\c  \prod_{j=1}^k S'_j$. If the operator
given by the integral kernel
$\gamma^{(k)}(\bold x_k, \bold x'_k)$ is positive (as an operator), then so is
 $S^{(k)}\gamma^{(k)}(\bold x_k,
\bold x_k')$, and the trace norm of $S^{(k)}\gamma^{(k)}$ is 
  \beaa
\|\gamma^{(k)}\|_{\mathcal{H}_k}
=  \int \big( S^{(k)}\gamma^{(k)}(\bold x_k,
\bold x_k')\big)\big|_{\x_k' =\x_k}\, d \bold x_k.
\eeaa

The authors of \cite{ESY1} prove  uniqueness of solutions to \eqref{G-P}
in the set of symmetric, positive  operators  $\ga_k$  satisfying, for some $C>0$,
\begin{equation}\label{theirspace}
\sup_{0 \le t \le T} \|\gamma^{(k)}
(t, \cdot, \cdot)\|_{\mathcal{H}_k} \le C^k
\end{equation}

 In that work, the equations \eqref{G-P} 
are obtained as a limit of the BBGKY hierarchy
(see \cite{ESY1}), 
and it is proved
that solutions to BBGKY  with initial conditions \eqref{special-IC}
 converge, in a weak sense, to
 a solution of \eqref{G-P} in the space \eqref{theirspace}.

The purpose of this note is to give a new proof of uniqueness of the Gross- Pitaevskii
hierarchy \eqref{G-P},  in a different space,  
motivated , in part, by  space-time   type  estimates, similar
in spirit to those
of \cite{KM}.

Our norms will be
\begin{equation} \label{newH}
 \|R^{(k)}\gamma^{(k)}(t, \cdot, \cdot)\|_{L^2 (\Bbb R^{3k} \times
\Bbb R^{3k})}
\end{equation} 
Here,
$R_j=(-\Delta_{x_j})^{1/2}$,
$R'_j=(-\Delta_{x'_j})^{1/2}$ 
and $R^{(k)}= \prod_{j=1}^k R_j \c \prod_{j=1}^k R'_j$. 
Notice that for a symmetric, 
smooth kernel $\gamma$, for which the associated linear
operator is positive we have
\beaa
 &&\|R^{(k)}\gamma^{(k)}(t, \cdot, \cdot)\|_{L^2 (\Bbb R^{3k} \times
\Bbb R^{3k})}\\
&\le& \|S^{(k)}\gamma^{(k)}(t, \cdot, \cdot)\|_{L^2 (\Bbb R^{3k} \times
\Bbb R^{3k})}\\
&\le&
\|\gamma^{(k)}(t, \cdot, \cdot)\|_{ \mathcal{H}_k} 
\eeaa
since $|S^{(k)}\gamma^{(k)} (\bold x, \bold x')|^2 \le S^{(k)}
\gamma^{(k)} (\bold x, \bold x) 
S^{(k)}\gamma^{(k)} (\bold x', \bold x')$. 
This is similar to the condition $a_{ii} a_{jj}-|a_{ij}|^2 \ge 0$
which is satisfied by all $n \times n$ 
positive semi-definite Hermitian matrices.

Our main result  is the following:

\begin{theorem}[Main Theorem] \label{maintheorem}
Consider  solutions  $\gamma^{(k)}(t, \bold x_k, \bold x'_k)$ 
of the Gross-Pitaevskii hierarchy \eqref{G-P}, with zero initial conditions,
which verify the estimates,
\bea\label{spacetime}
\int_0^T \|R^{(k)}B_{j, k}
\gamma^{(k)}(t, \cdot, \cdot)\|_{L^2 (\Bbb R^{3k} \times
\Bbb R^{3k})}
\, dt \le C^k
\eea
for some $C>0$ and all $1 \le j < k$. Then $
\|R^{(k)}\gamma^{(k)}(t, \cdot, \cdot)\|_{L^2 (\Bbb R^{3k} \times
\Bbb R^{3k})}=0
$ for all $k$ and all $t$.
\end{theorem}
Therefore,  solutions to \eqref{G-P} verifying the initial conditions
\eqref{special-IC}, are unique in
 the space-time norm \eqref{spacetime}. We plan to address the 
 connection with solutions of BBGKY in a future paper.
The following remark is  however reassuring.
\begin{remark} The sequence  $\gamma^{(k)}$,
  given by \eqref{eq:special-sol} with  $\phi$ 
an arbitrary   solution of 
 \eqref{eq:NLS} with $H^1$ data, 
 verifies  \eqref{spacetime} for  every 
 $T>0$ sufficiently small. Moreover, if
 the $H^1$ norm of the initial data is 
 sufficiently small then  \eqref{spacetime}
 is verified for all values of $T>0$.
 \end{remark}
 \begin{proof} 
 Observe that 
 $R^{(k)}B_{1, k}\gamma^{(k)}(t, x_1, \ldots, x_k; x_1' \ldots, x_k')$ can be written in the form,
\beaa
R^{(k)}B_{1, k}\gamma^{(k)}(t,\cdot,\cdot)
&=&R_1\left(|\phi(t, x_1)|^2 \phi(t, x_1)\right) R_2(\phi(t, x_2) \cdots R_k(\phi(t, x_k)\\
&\cdot&R'_1(\phi(t, x'_1) \cdots R'_k(\phi(t, x'_k)
\eeaa
Therefore, in $[0,T]\times \RRR^{3k}\times\RRR^{3k}$
we derive
\beaa
 \|R^{(k)}B_{j, k}
\gamma^{(k)}\|_{ L_t^1L^2 }&\le&  \|| R_1(|\phi|^2 \phi)\|_{L_t^1L_x^2}\cdot 
 \| R_2 \phi\|_{L_t^\infty L_x^2}\cdots \| R_k \phi\|_{L_t^\infty L_x^2}\\
 &\cdot&  \| R_1' \phi\|_{L_t^\infty L_x^2} \| R_2' \phi\|_{L_t^\infty L_x^2}\cdots \| R_k' \phi\|_{L_t^\infty L_x^2}\\
 &\le& C\|\nab(|\phi|^2\phi)\|_{L_t^1L_x^2}\times\|\nab \phi\|_{L_t^\infty L_x^2}^{2k-1}
\eeaa
where the norm on the left is 
in $ [0,T]\times \RRR^{3k}\times\RRR^{3k} $ and  all norms on the right hand side are taken relative to the space-time domain
$[0,T]\times \RRR^3$.

 In view of the  standard energy identity 
  for the nonlinear equation  \eqref{eq:NLS}  we 
 have    apriori  bounds for 
  $\sup_{t\in[0,T]}\|\nab \phi(t)\|_{L^2(\RRR^3)}$. 
  Therefore we only need to provide
   a  uniform  bound for the norm $\|\nab(|\phi|^2\phi)\|_{L_t^1L_x^2}$. We shall show below that this
   is possible for all values of $T>0$ provided
   that the $H^1$ norm of $\phi(0)$ is sufficiently
   small. The case of arbitrary size for $\|\phi(0)\|_{H^1}$ and sufficiently small $T$ is  easier and
   can be proved in a similar manner.

   We  shall  rely on  the following 
     Strichartz estimate (see \cite{KT}) for the linear, inhomogeneous,
     Schr\"odinger equation
  $i\partial_t \phi+\De\phi =f$ in $[0,T]\times\RRR^3$,
 \bea
  \|\phi\|_{L_t^2L_x^6}&\le& C\big( \| f\|_{L_t^1L_x^2}+
 \| \phi\|_{L_t^\infty L_x^2}\big)\label{Strich}
 \eea

  We start by using H\"older inequality, in $[0,T]\times\RRR^3$,
 \beaa
 \|\nabla( |\phi|^2 \phi)\|_{L_t^1 L_x^2 }&\le& C
 \|\nab\phi\|_{L_t^2 L_x^6}\|\phi^2\|_{L_t^2L_x^3}
 \le   C \|\nab\phi\|_{L_t^2  L_x^6}\|\phi\|_{L_t^4L_x^6}^2
 \eeaa
 Using \eqref{Strich} for $f=|\phi|^2\phi$ we derive,
 \beaa
  \|\nab \phi\|_{L_t^2L_x^6}&\le& C\big( 
  \| \nab (|\phi|^2\phi)\|_{L_t^1L_x^2}+
 \| \nab \phi\|_{L_t^\infty L_x^2}\big)\\
 \eeaa
 Denoting, 
  \beaa 
 A(T)&=&\|\, |\phi|^2 \phi\|_{L_t^1L_x^2([0,T]\times\RRR^3 )}\\
 B(T)&=&\|\nabla( |\phi|^2 \phi)\|_{L_t^1L_x^2([0,T]\times\RRR^3 )}
 \eeaa
 we derive,
 \beaa
 B(T)&\le& C\big(B(T)+\|\nab\phi(0)\|_{L^2}\big)
 \|\phi\|_{L_t^4L_x^6}^2 \\
 &\le&C\big(B(T)+\|\nab\phi(0)\|_{L^2}\big)\|\phi\|_{L_t^\infty L_x^6}\|\phi\|_{L_t^2L_x^6}\\
 &\le&C\big(A(T) + \|\phi(0)\|_{L^2}\big)\big(B(T)+\|\nab\phi(0)\|_{L^2}\big)
 \|\nab\phi(0)\|_{ L_x^2}
 \eeaa
 On the other hand, using \eqref{Strich} again,
 \beaa
 A(T)&\le& C\big( \|\phi^3\|_{L_t^1L_x^2}+\|\phi(0)\|_{L^2}\big)\\
 &\le&C(A(T)^3+\|\phi(0)\|_{L^2}\big)
 \eeaa
 Observe this last inequality  implies 
 that, for sufficiently  small  $\|\phi(0)\|_{L^2}$,  $A(T) $ remains
 uniformly bounded for all values of $T$.
 Thus, for all values of $T$, with another
 value of $C$,
 \beaa
 B(T)\le C\big(B(T)+\|\nab\phi(0)\|_{L^2}\big)
 \|\nab\phi(0)\|_{ L_x^2}
 \eeaa
 from which we get a uniform bound for $B(T)$
 provided that $\|\nab\phi(0)\|_{L^2}$ is  also sufficiently small.

\end{proof}

The proof of Theorem \eqref{maintheorem} is based on two ingredients. 
One is expressing $\gamma^{(k)}$
in terms of the {\it  future iterates}  $\gamma^{(k+1)} \cdots$, $\gamma^{(k+n)}$
using Duhamel's formula. 
Since $B^{(k+1)} =\sum_{j=1}^{k}B_{j, \, k+1}$ 
is a sum 
of $k$ terms, the iterated Duhamel's formula involves $k (k+1) \cdots (k+n-1)$
terms. These have to be grouped into much fewer $O(C^n)$ sets of terms. This
part of our paper follows in  the spirit of  the Feynman path  combinatorial 
arguments of \cite{ESY1}. The second ingredient is the main
  novelty of our work. We derive   a space-time  estimate, reminiscent of
   the bilinear estimates of \cite{KM}. 

\begin{theorem}\label{ourest}

Let $\gamma^{(k+1)}(t, \bold x_{k+1}, \bold x'_{k+1}) $ verify 
the  homogeneous equation,
\bea
&&\left( i \partial_t +\De_{\pm}^{k+1}\right) 
\gamma^{(k+1)}  =0,\qquad \De_{\pm}^{(k+1)}= \Delta_{\bold x_{k+1}} - \Delta_{\bold x'_{k+1}} \label{mainest}
\\
&& \gamma^{(k+1)}(0, \bold x_{k+1}, \bold x'_{k+1})=\gamma_0^{(k+1)}(\bold x_{k+1}, \bold x'_{k+1}). \nn
\eea
Then there exists a constant $C$, independent of $j, k$, such that
\bea
&\|R^{(k)}B_{j, k+1}(\gamma^{(k+1)}) \label{eq:homog}
\|_{L^2(\Bbb R \times \Bbb R^{3k} \times \Bbb R^{3k})}\\
&\le C \notag
\|R^{(k+1)}\gamma_0^{(k+1)}\|_{L^2(\Bbb R^{3(k+1)} \times \Bbb R^{3(k+1)})}
\eea

\end{theorem}

 \section{Proof of the estimate}
Without loss of generality, we may take  $j=1$ in $B_{j, k+1}$. It also suffices  to  estimate  the term in $B^1_{j, k+1}$, the term in $B^2_{j, k+1}$  can
 be treated in the same manner.
Let $\gamma^{(k+1)}$ be as in \eqref{mainest}. Then the 
Fourier transform of 
$\gamma^{(k+1)}$  with respect to the variables $(t, \bold x_k , \bold x'_k)$
is given by the formula,
\beaa
\delta(\tau + |\xib_k|^2 - |\xib_k'|^2) {\hat \gamma}(\xi, \xi')
\eeaa
where $\tau $ corresponds to the time  $t$ and 
$\xib_k=(\xi_1,\xi_2,\ldots, \xi_k)$, 
$\xib_k'=(\xi_1',\xi_2',\ldots, \xi_k')$ correspond to the space variables $\bold x_k=(x_1,x_2,\ldots, x_k)$
and  $\bold x_k'=(x_1',x_2',\ldots, x_k')$.  
We also write $\xib_{k+1}=(\xib_k, \xi_{k+1})$,
$\xib_{k+1}'=(\xib_k', \xi_{k+1}')$ and,
\beaa
|\xib_{k+1}|^2&=&|\xi_1|^2+\ldots+  |\xi_k|^2+ | \xi_{k+1}|^2=|\xib_k|^2+ | \xi_{k+1}|^2\\
|\xib_{k+1}'|^2&=&|\xi_1'|^2+\ldots+  |\xi'_k|^2+ | \xi_{k+1}'|^2
\eeaa
The Fourier transform of
$B^1_{1, k+1}(\gamma^{(k+1)})$, with respect to the same variables
 $(t, \bold x_k , \bold x'_k)$, 
is given by,
\bea
\int \int\delta(\cdots){\hat \gamma}(\xi_1-\xi_{k+1}-\xi'_{k+1}, \xi_2, \cdots , \xi_{k+1},
\xib'_{k+1})
d \xi_{k+1} 
\, d \xi'_{k+1}\qquad\,\,   
 \label{FT}
\eea
where, 
\beaa
\de(\cdots)&=& \delta( \tau +|\xi_1-\xi_{k+1}-\xi'_{k+1}|^2 +
|\xib_{k+1}|^2 -|\xi_1|^2
-|\xib'_{k+1}|^2)
\eeaa
and  $\gamma$ denotes the initial condition $ \gamma_0^{(k+1)}$.
By Plancherel's theorem,  estimate 
\eqref{eq:homog}  is equivalent to 
the following estimate,
\bea
\|I_k[ f]\|_{L^2(\RRR \times \RRR^k \times \RRR^k)}\le
C \|{\hat f}\|_{L^2( \RRR^{k+1} \times \RRR^{k+1})},
\eea
applied to  $f=R^{(k+1)}\gamma$,
where,
\bea
I_k[ f](\tau,\xib_k, \xib'_k)=\int\int \delta( \ldots)
\frac{|\xi_1| {\hat f}(\xi_1-\xi_{k+1}-\xi'_{k+1}, \xi_2, \cdots , \xi_{k+1},
 \xib'_{k+1})}{
|\xi_1-\xi_{k+1}-\xi'_{k+1}| |\xi_{k+1}||\xi'_{k+1}|}
d \xi_{k+1} 
\, d \xi'_{k+1}.\nn
\label{eq:I}
\eea

Applying the Cauchy-Schwarz inequality with measures,
we easily check that,
\beaa
|I_k[ f]|^2&\le&\int\int \delta( \ldots)
\frac{|\xi_1|^2 }{
|\xi_1-\xi_{k+1}-\xi'_{k+1}|^2 |\xi_{k+1}|^2|\xi'_{k+1}|^2}
d \xi_{k+1} 
\, d \xi'_{k+1}\\
&\c&\int\int \delta( \ldots)| {\hat f}(\xi_1-\xi_{k+1}-\xi'_{k+1}, \xi_2, \cdots , \xi_{k+1},
 \xib'_{k+1})|^2d \xi_{k+1} 
\, d \xi'_{k+1}
\eeaa
If we can show that the supremum over $\tau$,
$\xi_1
\cdots  \xi_k, \xi'_1
\cdots  \xi'_k$
of the first integral above
 is bounded by a constant $C^2$, we
 infer that,
 \beaa
\|I_k[ f]\|_{L^2}^2&\le& C^2
\int \int \int \delta( \ldots)| {\hat f}(\xi_1-\xi_{k+1}-\xi'_{k+1}, \xi_2, \cdots , \xi_{k+1},
 \xib'_{k+1})|^2d \xib_{k+1} d \xib'_{k+1}
  d\tau \\
&  \le &
C^2 \|{\hat f}\|_{L^2( \RRR^{k+1} \times \RRR^{k+1})}^2
\, \eeaa
 Thus,
\beaa
\|I_k[ f]\|_{L^2(\RRR \times \RRR^k \times \RRR^k)}^2\le C^2 
 \|{\hat f}\|_{L^2( \RRR^{k+1} \times \RRR^{k+1})}^2
\eeaa
as desired.
Thus we have reduced matters to the following,

\begin{proposition} \label{measureprop}
There exists a constant $C$ such that
\beaa
\int&& \delta( \tau +|\xi_1-\xi_{k+1}-\xi'_{k+1}|^2 
+|\xi_{k+1}|^2 
-|\xi'_{k+1}|^2)\\
&&\frac{|\xi_1|^2}{
|\xi_1-\xi_{k+1}-\xi'_{k+1}|^2 |\xi_{k+1}|^2|\xi'_{k+1}|^2}
d \xi_{k+1} d \xi'_{k+1} \notag
\le C
\eeaa
uniformly in $\tau, \xi_1$.
\end{proposition}

The proof is based on the following lemmas

\begin{lemma}\label{lemma}
Let $P$ be a 2 dimensional plane or sphere in $\RRR^3$ with the usual
induced surface measure $dS$. Let $0<a<2$, $0<b<2$, $a+b>2$. Let $\xi\in \RRR^3$. Then there exists $C$ independent of $\xi$ and $P$ such that

\[
\int_P \frac{1}{|\xi - \eta|^a |\eta|^b} dS(\eta)
\le \frac{C}{|\xi|^{a+b-2}}
\]
\end{lemma}

\begin{proof} If $P=\RRR^2$ and $\xi \in \RRR^2$ this is well known.
The same proof works in our case, by breaking up the integral $I \le I_1 + I_2 + I_3$ over the
overlapping regions: 

Region 1:
$\frac{|\xi|}{2} < |\eta| < 2 |\xi|$.
In this region
$|\xi - \eta| < 3 |\xi|$ and
\begin{align}
I_1 \le &C \frac{1}{|\xi|^b} \notag
\int_{P \cap \{|\xi - \eta| < 3 |\xi|\}} \frac{1}{|\xi - \eta|^a} dS(\eta)\\
\le &C  \frac{1}{|\xi|^b} \sum_{i = - \infty}^{1} \notag
\int_{P \cap \{3^{i-1} |\xi| < |\xi - \eta| < 3^i |\xi|\}} \frac{1}{|\xi - \eta|^a} dS(\eta)\\
\le &C  \frac{1}{|\xi|^b} 
\sum_{i = - \infty}^{1} \notag
\frac{1}{|3^i \xi|^a} (3^i |\xi|)^2 =\frac{C}{|\xi|^{a+b-2}}
\end{align}
 where we have used the obvious fact that the area of
$ P \cap \{3^{i-1} |\xi| < |\xi - \eta| < 3^i |\xi|\}$ is $\le C
 (3^i |\xi|)^2$.

Region 2:
$\frac{|\xi|}{2} < |\xi - \eta| < 2 |\xi|$. In this region
$|\eta| < 3 |\xi|$ and

\begin{align}
I_2 \le &C \frac{1}{|\xi|^a} \notag
\int_{P \cap \{|\eta| < 3 |\xi|\}} \frac{1}{|\eta|^b} dS(\eta)\\
& \le\frac{C}{|\xi|^{a+b-2}}
\end{align}
in complete analogy with region 1.

Region 3:  $|\eta| > 2 |\xi|$ 
or   $|\xi -\eta| > 2 |\xi|$.
In this region, $|\eta| >  |\xi|$ and 
 $2 |\xi - \eta| \ge |\xi - \eta| + |\xi| \ge |\eta|$, thus
\begin{align}
I_3 \le &\notag C
\int_{P \cap \{| \eta| >  |\xi|\}} \frac{1}{|\eta|^{a+b}} dS(\eta)\\
\le &C   \sum_{i = 1}^{\infty} \notag
\int_{P \cap \{2^{i-1} |\xi| < |\eta| < 2^i |\xi|\}} 
\frac{1}{| \eta|^{a+b}} dS(\eta)\\
\le &C   
\sum_{i = 1}^{\infty} \notag
\frac{1}{|2^i \xi|^{a+b}} (2^i |\xi|)^2 =\frac{C}{|\xi|^{a+b-2}}\\
\end{align}
\end{proof}

We also have
\begin{lemma}
Let $P$, $\xi$ as in Lemma \eqref{lemma} and let $\epsilon = \frac{1}{10}$.
Then
\[
\int_P \frac{1}{|\xi - \eta|^{2- \epsilon}
|\frac{\xi}{2} - \eta|
 |\eta|^{2 - \epsilon}} dS(\eta)
\le \frac{C}{|\xi|^{3 - 2 \epsilon}}
\]
\end{lemma}

\begin{proof}
Consider separately the regions $|\eta|> \frac{|\xi|}{2}$,
and $|\xi - \eta|> \frac{|\xi|}{2}$,
and apply the previous lemma.\end{proof}

We are ready to prove the main estimate
of Proposition  \eqref{measureprop}

\begin{proof}
Changing $k+1$ to 2, we have to show

\begin{align}
I=\int \delta( \tau +|\xi_1-\xi_2-\xi'_2|^2 
+|\xi_2|^2 \notag
-|\xi'_2|^2)
\frac{|\xi_1|^2}{
|\xi_1-\xi_2-\xi'_2|^2 |\xi_2|^2|\xi'_2|^2}
d \xi_2 d \xi'_2 \notag
\le C
\end{align}

The integral is symmetric in $ \xi_1 - \xi_2 - \xi'_2$ and $\xi_2$
so we can integrate, without loss of generality, over
$| \xi_1 - \xi_2 - \xi'_2|>|\xi_2|$.

Case 1. Consider the integral $I_1$ restricted to the region $|\xi'_2|> |\xi_2|$
and integrate $\xi'_2$ first.

Notice
\begin{align}
&\delta( \tau +|\xi_1-\xi_2-\xi'_2|^2 
+|\xi_2|^2 \notag
-|\xi'_2|^2) d \xi'_2\\
&=\delta( \tau +|\xi_1-\xi_2|^2 -2 (\xi_1 - \xi_2)\cdot \xi'_2 
+|\xi_2|^2 \notag) d \xi'_2\\
&=\frac{dS(\xi'_2)}{2 |\xi_1- \xi_2|}
\end{align}

where $dS$ is surface measure on a plane $P$ in $\RRR^3$, i.e. the plane
$\xi'\cdot \om=\la$ with $\om\in \SSS^2$ and
  $\la=\frac{ \tau +|\xi_1-\xi_2|^2+|\xi_2|^2}{ 2 |\xi_1- \xi_2|}$.

In this region

\begin{align}
I_1 &\le |\xi_1|^2 \int_{\RRR^3} \frac{d \xi_2}{|\xi_2|^2|\xi_1-\xi_2|} \notag
\int_P \frac{dS(\xi'_2)}{|\xi_1- \xi_2 - \xi'_2|^2|\xi'_2|^2}\\
& \le |\xi_1|^2 \int_{\RRR^3} \notag
\frac{d \xi_2}{|\xi_2|^{2+ 2 \epsilon}
|\xi_1-\xi_2|}
\int_P \frac{dS(\xi'_2)}{|\xi_1- \xi_2 - \xi'_2|^{2- \epsilon}|\xi'_2|^{2- \epsilon}}\\
&\le C |\xi_1|^2 \int_{\RRR^3} \notag
\frac{d \xi_2}{|\xi_2|^{2+ 2 \epsilon}
|\xi_1-\xi_2|^{3 - 2 \epsilon}}\\
&\le C
\end{align}

Case 2. Consider the integral $I_2$ restricted to the region $|\xi'_2|< |\xi_2|$
and integrate $\xi_2$ first.

Notice
\begin{align}
&\delta( \tau +|\xi_1-\xi_2-\xi'_2|^2 
+|\xi_2|^2 \notag
-|\xi'_2|^2) d \xi_2\\
\,\,\,&=\delta\left( \tau +|\frac{\xi_1-\xi'_2}{2}- (\xi_2 - \frac{\xi_1-\xi'_2}{2})
|^2
+|(\xi_2 - \frac{\xi_1-\xi'_2}{2}) + \frac{\xi_1-\xi'_2}{2}|^2
- |\xi'_2|^2
 \notag \right) d \xi_2\\
&= \delta \left( \tau + \frac{|\xi_1 - \xi'_2|^2}{2}
+ 2 |\xi_2 - \frac{\xi_1-\xi'_2}{2}|^2 - |\xi'_2|^2 \right) d \xi_2\\
&= \notag \frac{dS(\xi_2)}{4 |\xi_2- \frac{\xi_1-\xi'_2}{2}|}
\end{align}

where $dS$ is surface measure on a sphere $P$, i.e. the sphere
 in $\xi_2$ centered at  $\frac{1}{2}(\xi_1-\xi_2')$ and radius 
 $\frac{1}{2}\big( |\xi'_2|^2 -\tau- \frac{|\xi_1 - \xi'_2|^2}{2}\big)$.

Thus

\begin{align}
I_2 &\le |\xi_1|^2 \int_{\RRR^3} \frac{d \xi'_2}{|\xi'_2|^2}\notag
\int_P \frac{dS(\xi_2)}{|\xi_2|^2|\xi_2 - \frac{\xi_1-\xi'_2}{2}|
|\xi_1 - \xi_2 - \xi'_2|^2}\\
&\le  |\xi_1|^2 \int_{\RRR^3} \frac{d \xi'_2}{|\xi'_2|^{2+2 \epsilon}}\notag
\int_P \frac{dS(\xi_2)}{|\xi_2|^{2- \epsilon}|\xi_2 - \frac{\xi_1-\xi'_2}{2}|
|\xi_1 - \xi_2 - \xi'_2|^{2- \epsilon}}\\
&\le  |\xi_1|^2 \int_{\RRR^3} \frac{d \xi'_2}{|\xi'_2|^{2+2 \epsilon}|
\xi_1 - \xi'_2|^{3 - 2 \epsilon}}\notag
\le C
\end{align}
\end{proof}

\section{Duhamel expansions and regrouping}

This part of our note is based on a somewhat
  shorter  version of the combinatorial
ideas  of  \cite{ESY1}. We are 
grateful to  Schlein and Yau for explaining their 
arguments  to us.

Recalling the notation  $\Delta^{(k)}_{\pm}= \Delta_{\bold x_k} - 
\Delta_{\bold x'_k}$
and $\Delta_{\pm, x_j}= \Delta_{x_j}- \Delta_{x'_j}$ we write,
\bea
 \gamma^{(1)}(t_1, \cdot)&=&
\int_0^{t_1} e^{i(t_1-t_2) \Delta^{(1)}_{\pm}}B_{2}(\gamma^2
(t_2)) \, dt_2\notag\\
&=&\int_0^{t_1} \int_0^{t_2} e^{i(t_1-t_2) \Delta^{(1)}_{\pm}}B_{2}
 e^{i(t_2-t_3) \Delta^{(2)}_{\pm}}
(\gamma^3
(t_3)) \, dt_2dt_3\notag\\
&=&\cdots\cdots \cdots\cdots\nn\\
&=&\int_0^{t_1}\cdots \int_0^{t_n}J(\tb_{n+1})
\, dt_2 \,\cdots  dt_{n+1}\label{duhamel}
\eea
where, $\tb_{n+1}=(t_1,\ldots, t_{n+1})$
and 
\beaa
J( \tb_{n+1})&=& e^{i(t_1-t_2) \Delta^{(1)}_{\pm}}
B_2
\cdots e^{i(t_n-t_{n+1}) \Delta^{(n)}_{\pm}}B_{n+1}
(\gamma^{(n+1)})(t_{n+1}, \cdot)
\eeaa
Expressing
 $B^{(k+1)} =\sum_{j=1}^{k}B_{j, \, k+1}$, $\cdots$, 
the integrand $J(\tb_{n+1})=J(t_1,\ldots, t_{n+1})$
in \eqref{duhamel} can be written as
\bea
J(\tb_{n+1})&=&\sum_{\mu \in M}J( \tb_{n+1};\, \mu)
\eea
where,
\beaa
 J(\tb_{n+1};\, \mu)
&=&e^{i(t_1-t_2) \Delta^{(1)}_{\pm}}
B_{1,2}
e^{i(t_2-t_3) \Delta^{(2)}_{\pm}}
B_{\mu(3), 3}
\cdots\\
&& e^{i(t_n-t_{n+1}) \Delta^{(n)}_{\pm}}B_{\mu (n+1), n+1}
(\gamma^{(n+1)})(t_{n+1}, \cdot)\nn
\eeaa
Here we  have  denoted by $M$ the set of  maps 
$\mu: \{2, \cdots, n+1\} \to \{1, \cdots n \}$
satisfying $\mu(2)=1$ and  $\mu(j) < j$ for all $j$. 

Graphically, such a $\mu$ can be represented by selecting one $\bold{B}$ entry
from each column of an $n \times n$ matrix

such as, for example, (if $\mu(2)=1$, $\mu(3)=2$, $\mu(4)=1$, etc),
\begin{equation} \label{matrix}
\begin{pmatrix} 
\bold{B_{1,2}}&B_{1,3}&\bold{B_{1,4}}& \cdots & \bold{B_{1, n+1}}\\
0&\bold{B_{2, 3}}&B_{2, 4}& \cdots & B_{2, n+1} \\
0&0 & B_{3, 4}& \cdots & B_{3, n+1} \\
\cdots & \cdots & \cdots & \cdots & \cdots\\
0 & 0 & 0 & \cdots & B_{n, n+1}\\
\end{pmatrix}
\end{equation}

To such a matrix one can associate a Feynman  graph whose nodes are the selected entries, as in  \cite{ESY1}. However, our exposition will be
self-contained and will   not  rely  explicitly 
on the Feynman  graphs.

We will   consider
\begin{equation}
\II(\mu, \sigma)
=\int_{t_1 \ge t_{\sigma(2)} \ge t_{\sigma(3)} \ge \cdots \ge t_{\sigma(n+1)}} \label{sigmamu}
 J( \tb_{n+1};\, \mu)
dt_2 \cdots dt_{n+1}
\end{equation}
where $\sigma$ is a permutation of $\{2, \cdots , n+1\}$.  The integral $\II(\mu,\si)$  is represented by $(\mu, \sigma)$, or, graphically,  by the matrix
\begin{equation} \label{matrixsigma}
\begin{pmatrix} 
t_{\sigma^{-1} (2)}&t_{\sigma^{-1}(3)}&t_{\sigma^{-1}(4)}& \cdots & t_{\sigma^{-1}(n+1)}\\
\bold{B_{1,2}}&B_{1,3}&\bold{B_{1,4}}& \cdots & \bold{B_{1, n+1}}&
\mbox{row 1}\\
0&\bold{B_{2, 3}}&B_{2, 4}& \cdots & B_{2, n+1} & \mbox{row 2}\\
0&0 & B_{3, 4}& \cdots & B_{3, n+1}& \mbox{row 3}\\
\cdots & \cdots & \cdots & \cdots & \cdots\\
0 & 0 & 0 & \cdots & B_{n, n+1} & \mbox{row n}\\
\mbox{column 2} & \mbox{column 3} & \mbox{column 4} & \cdots &
\mbox{column n+1}\\
\end{pmatrix}
\end{equation}

Notice the columns of this matrix are labelled $2 $ to $n+1$, while the
rows are labelled $1$ through $n$.

We will define a set of ``acceptable moves'' on the
set of such matrices.
 Imagine a board game where the
names $B_{i, j}$ 
are carved in, and one entry $B_{\mu(j), j}$, $\mu(j)<j$,
in each column is highlighted.
If $\mu(j+1)< \mu(j)$, the player is allowed
to exchange the highlighted entries in columns
$j$ and $j+1$ and, at the same time, exchange the highlighted
entries in rows $j$ and $j+1$. 
This changes $\mu$ to a new $\mu'=(j, j+1)\circ \mu \circ (j, j+1)$. 
Here $(j, j+1)$  denotes the permutation which reverses $j$ and $j+1$.
The rule for changing $\sigma$
is $\sigma'^{-1}= \sigma^{-1}\circ (j, j+1)$.
In other words, $\sigma^{-1}$ changes according to column exchanges.
Thus going from

\begin{equation}
\begin{pmatrix} \label{equivmatrix1}
t_2&t_5&t_4 & t_3\\
\bold{B_{1,2}}&B_{1,3}&\bold{B_{1,4}}&  B_{1, 5}\\
0&\bold{B_{2, 3}}&B_{2, 4} & B_{2, 5}\\
0&0 & B_{3, 4}& B_{3, 5}\\
0 & 0 & 0 &  \bold{B_{4, 5}}\\
\end{pmatrix}
\end{equation}
 
to

\begin{equation}
\begin{pmatrix} \label{equivmatrix2}
t_2&t_4&t_5 & t_3\\
\bold{B_{1,2}}&\bold{B_{1,3}}&B_{1,4}&  B_{1, 5}\\
0&B_{2, 3}&\bold{B_{2, 4}} & B_{2, 5}\\
0&0 & B_{3, 4}& \bold{B_{3, 5}}\\
0 & 0 & 0 &  B_{4, 5}\\
\end{pmatrix}
\end{equation}
is an acceptable move. In the language of \cite{ESY1} the partial
order of the graphs is preserved by an acceptable move.

The relevance  of this game  to our 
situation is explained by  the following,

\begin{lemma} \label{ints}
Let $(\mu, \sigma)$ 
be transformed into $(\mu', \sigma')$ by an acceptable move.
Then, for the corresponding integrals \eqref{sigmamu},
 $\II(\mu, \sigma) = \II(\mu', \sigma')$
\end{lemma}
\begin{proof}

We will first explain the strategy of the proof  by focusing on   an explicit  example.
Consider the integrals $\II_1$ corresponding to \eqref{equivmatrix1} and $\II_2$ corresponding to  \eqref{equivmatrix2}:

\bea
\II_1&=& \int_{t_1 \ge t_2\ge t_5 \ge t_4 \ge t_3 } e^{i(t_1-t_2)\Delta^{(1)}_{\pm}}
B_{1, 2}\label{ex1}
e^{i(t_2-t_3) \Delta^{(2)}_{\pm}}B_{2, 3}\\
&\quad& e^{i(t_3-t_4) \Delta^{(3)}_{\pm}}B_{1, 4}\notag
e^{i(t_4-t_5) \Delta^{(4)}_{\pm}}B_{4, 5}(\gamma^{(5)}(t_5, \cdot)
dt_2 \cdots dt_5\\
\II_2&=& \int_{t_1 \ge t_2\ge t_5 \ge t_3 \ge t_4 } e^{i(t_1-t_2)\Delta^{(1)}_{\pm}}
B_{1, 2}\label{ex2}
e^{i(t_2-t_3) \Delta^{(2)}_{\pm}}B_{1, 3}\\
&& e^{i(t_3-t_4) \Delta^{(3)}_{\pm}}B_{2, 4}
e^{i(t_4-t_5) \Delta^{(4)}_{\pm}}B_{3, 5}(\gamma^{(5)}(t_5, \cdot)
dt_2 \cdots dt_5\notag
\eea
We first observe    the identity, 
\bea
&&e^{i(t_2-t_3) \Delta^{(2)}_{\pm}}B_{2, 3}
e^{i(t_3-t_4) \Delta^{(3)}_{\pm}}B_{1, 4}\notag
e^{i(t_4-t_5) \Delta^{(4)}_{\pm}}\\
=&&e^{i(t_2-t_4) \Delta^{(2)}_{\pm}}B_{1, 4}
e^{-i(t_3-t_4)( \Delta^{(3)}_{\pm}
-\Delta_{\pm, x_3} + \Delta_{\pm, x_4})}
B_{2, 3}\notag
e^{i(t_3-t_5) \Delta^{(4)}_{\pm}}
\eea

In other words, $(t_3, x_3, x'_3)$ and $(t_4, x_4, x'_4)$ and the 
position of $B_{2, 3}$ and $B_{1, 4}$ have been exchanged.
This is based on trivial commutations, and is proved below in general,
see\eqref{comm}.

Recalling the definition \eqref{Bs} we abbreviate the integral
kernel of $B_{j, k+1}$,
$$ \delta_{j, k+1}=
\delta(x_j - x_{k+1}) \delta (x_j - x'_{k+1})
-\delta(x'_j - x_{k+1}) \delta (x'_j - x'_{k+1}). $$
We also denote,  
\beaa
\gamma_{3, 4}&=&\gamma^{(5)}(t_5, x_1, x_2, x_3, x_4, x_5 ;
x'_1, x'_2, x'_3, x'_4, x'_5)\\
\gamma_{4, 3}&=&\gamma^{(5)}(t_5, x_1, x_2, x_4, x_3, x_5 ;
x'_1, x'_2, x'_4, x'_3, x'_5).
\eeaa
By the symmetry assumption \eqref{symm}, $\gamma_{3, 4}=\gamma_{4, 3}$.
Thus the integral $\II_1$ of  \eqref{ex1} equals
\bea
\II_1&=& \int_{t_1 \ge t_2\ge t_5 \ge t_4 \ge t_3 } 
\int_{\bold R^{24}}
e^{i(t_1-t_2)\Delta^{(1)}_{\pm}}
\delta_{1, 2} \label{int11}\\
&\quad&e^{i(t_2-t_4) \Delta^{(2)}_{\pm}}\delta_{1, 4}
e^{-i(t_3-t_4)( \Delta^{(3)}_{\pm}
-\Delta_{\pm, x_3} + \Delta_{\pm, x_4})}
\delta_{2, 3} \notag
e^{i(t_3-t_5) \Delta^{(4)}_{\pm}}\\
&\quad&\delta_{4, 5}\gamma_{4, 3}
dt_2 \cdots dt_5 d x_2 \cdots d x_5 d x'_2 \cdots d x'_5 \notag
\eea

In the above integral we perform the  change of variables which  exchanges
$(t_3, x_3 , x'_3)$ with $(t_4, x_4 , x'_4)$.
Thus, in particular, $\Delta_{x_3}$ becomes $\Delta_{x_4}$,  and $\De_{\pm}^{(3)}=
\De_{\pm}^{(2)}+\De_{\pm, x_3}$ becomes
$\De_{\pm}^{(2)}+\De_{\pm, x_4}$. Thus,
$ \Delta^{(3)}_{\pm}
-\Delta_{\pm, x_3} + \Delta_{\pm, x_4}$
changes to,
\beaa
\De_{\pm}^{(2)}+\De_{\pm, x_4}-\Delta_{\pm, x_4} + \Delta_{\pm, x_3}=\De_{\pm}^{(3)},
\eeaa
and $\De_{\pm}^{(4)}$ stays unchanged.
The integral \eqref{int11}
becomes
\beaa
\II_1&=& \int_{t_1 \ge t_2\ge t_5 \ge t_3 \ge t_4 } 
\int_{\bold R^{24}}
e^{i(t_1-t_2)\Delta^{(1)}_{\pm}}
\delta_{1, 2}\\
&\quad &e^{i(t_2-t_3) \Delta^{(2)}_{\pm}}\delta_{1, 3}
e^{i(t_3-t_4) \Delta^{(3)}_{\pm}}\notag
\delta_{2, 4}
e^{i(t_4-t_5) \Delta^{(4)}_{\pm}}\\
&\quad&\delta_{3, 5}\gamma_{3, 4} \notag
dt_2 \cdots dt_5 d x_2 \cdots d x_5 d x'_2 \cdots d x'_5\\
&=& \int_{t_1 \ge t_2\ge t_5 \ge t_3 \ge t_4 } e^{i(t_1-t_2)\Delta^{(1)}_{\pm}}
B_{1, 2}
e^{i(t_2-t_3) \Delta^{(2)}_{\pm}}B_{1, 3}\\
&& e^{i(t_3-t_4) \Delta^{(3)}_{\pm}}B_{2, 4}\notag
e^{i(t_4-t_5) \Delta^{(4)}_{\pm}}B_{3, 5}(\gamma^{(5)}(t_5, \cdot)
dt_2 \cdots dt_5\\
&=&\II_2
\eeaa
Therefore,
 $\II_1=\II_2$ as stated above.

  Notice the domain of integration
corresponds to $\sigma'(2)=2$, $\sigma'(3)=5$,
$\sigma'(4)=3$, $\sigma'(5)=4$, that is, $\sigma'=(3, 4) \circ \sigma$.

Now we proceed to the general case.
Consider a  typical term,
 \bea
 \II(\mu, \sigma)&=&\int_{t_1\ge \cdots t_{\sigma (j)}\ge t_{\sigma (j+1)}\ge\cdots t_{\sigma(n+1)}\ge 0 } \label{firstint}
J( \tb_{n+1};\,\mu)dt_2\ldots dt_{n+1} \\
&=&\notag \int_{t_1\ge \cdots t_{\sigma (j)}\ge t_{\sigma (j+1)}\ge\cdots t_{\sigma(n+1)}\ge 0 }\\
&&\cdots e^{i(t_{j-1}-t_j) \Delta^{(j-1)}_{\pm}} 
B_{l, j}
e^{i(t_j-t_{j+1}) \Delta^{(j)}_{\pm}}
B_{i, j+1}
 e^{i(t_{j+1}-t_{j+2}) \Delta^{(j+1)}_{\pm}}\notag\\
&&(\cdots) dt_2\ldots dt_{n+1} 
\label{J}
 \eea
  with associated matrix of
the form,
\begin{equation}
\begin{pmatrix} \label{matrix1}
\cdots & t_{\sigma^{-1}(j)} & t_{\sigma^{-1}(j+1)} & \cdots \\
\cdots& B_{i, j}& \bold{B_{i, j+1}}&\cdots\\
\cdots& \cdots & \cdots & \cdots \\
\cdots& \bold{B_{l, j}}&  B_{l, j+1}&\cdots\\
\cdots& \cdots & \cdots & \cdots \\
\cdots& \cdots & \cdots & \mbox{row j} \\
\cdots& \cdots & \cdots & \mbox{row j+1} \\
\cdots& \cdots & \cdots & \cdots \\
\end{pmatrix}
\end{equation}
where    
$\mu(j)=l$ and  $\mu(j+1)=i$ and $i<l<j<j+1$.
It is understood that rows j or j+1 may in fact not have highlighted entries, as in the previous
example. We plan to show 

\bea
\II=\II' \label{I=I'}
\eea
where
 \bea
 \II'
&=& \int_{t_1\ge \cdots t_{\sigma' (j)}\ge t_{\sigma' (j+1)}\ge\cdots \notag
t_{\sigma'(n+1)}\ge 0 }\\\label{J'}
&&\cdots e^{i(t_{j-1}-t_j) \Delta^{(j-1)}_{\pm}}
B_{i, j}
e^{i(t_j-t_{j+1}) \Delta^{(j)}_{\pm}}
B_{l, j+1}
 e^{i(t_{j+1}-t_{j+2}) \Delta^{(j+1)}_{\pm}}\notag\\
&&(\cdots)'dt_2\ldots dt_{n+1} 
\label{J'}
 \eea
The $\cdots$ at the beginning of \eqref{J} and \eqref{J'} are the same.
Any $B_{j, \alpha}$ 
in $(\cdots)$ in \eqref{J} become
$B_{j+1, \alpha}$  in $(\cdots)'$ in \eqref{J'}.
Similarly, any $B_{j+1, \alpha}$ 
in $(\cdots)$ in \eqref{J} become
$B_{j, \alpha}$ in  $(\cdots)'$ in  \eqref{J'}, while the
rest is unchanged.

Thus $\II'$ is represented by the matrix,
 \begin{equation}
\begin{pmatrix}
\cdots& t_{\sigma'^{-1}(j)} & t_{\sigma'^{-1}(j+1)} & \cdots \\ \label{matrix2}
\cdots& \bold{B_{i, j}}& B_{i, j+1}&\cdots\\
\cdots& \cdots & \cdots & \cdots \\
\cdots& B_{l, j}&  \bold{B_{l, j+1}}&\cdots\\
\cdots& \cdots & \cdots & \cdots \\
\cdots& \cdots & \cdots & \mbox{(row j)'} \\
\cdots& \cdots & \cdots & \mbox{(row j+1)'} \\
\cdots& \cdots & \cdots & \cdots \\
\end{pmatrix}
\end{equation}
where the highlighted entries of (row j)', respectively (row j+1)'
in \eqref{matrix2} have the positions of the highligted
entries of row j+1 , respectively row j
in \eqref{matrix1}.

To prove \eqref{I=I'}
denote, $\tilde{\Delta}_{\pm}^{(j)}
= \Delta_{\pm}^{(j)} -\Delta_{\pm, x_j} +\Delta_{\pm, x_{j+1}}$.
We consider  the terms,
\bea
P=
B_{l, j}
e^{i(t_j-t_{j+1}) \Delta^{(j)}_{\pm}}
B_{i, j+1}
 \label{first}
\eea
and 
\bea
 \tilde P =
B_{i, j+1}
e^{-i(t_j-t_{j+1}) \tilde{\Delta}^{(j)}_{\pm}}
B_{l, j}
 \label{second}
\eea
We will show that,
\bea
e^{i(t_{j-1}-t_j) \Delta^{(j-1)}_{\pm}} P e^{i(t_{j+1}-t_{j+2}) \Delta^{(j+1)}_{\pm}} \label{comm}
=e^{i(t_{j-1}-t_{j+1})\Delta^{(j-1)}_{\pm}}\tilde P  e^{i(t_{j}-t_{j+2}) \Delta^{(j+1)}_{\pm}}
\eea

Indeed  in \eqref{first}  we can write  
$\Delta_{\pm}^{(j)}=\Delta_{\pm, x_i} + 
( \Delta_{\pm}^{(j)}-\Delta_{\pm, x_i} )$. Therefore,
\beaa
e^{i(t_j-t_{j+1})\Delta_{\pm}^{(j)}}=e^{i(t_j-t_{j+1})\Delta_{\pm, x_i}} \,\c\, 
e^{i(t_j-t_{j+1})( \Delta_{\pm}^{(j)}-\Delta_{\pm, x_i} )}
\eeaa
 Observe that the first terms on the right  can be commuted to the left of $B_{l, j}$,
the second one to the right of $B_{i, j+1}$ in the expression for $I$.
Thus,
\beaa
P&=& 
e^{i(t_j-t_{j+1})(\Delta_{\pm, x_i})}
 B_{l, j} B_{i, j+1}
 e^{i(t_j-t_{j+1})(\Delta^{(j)}_{\pm}-\Delta_{\pm, x_i})}
\eeaa
and
\beaa
&&e^{i(t_{j-1}-t_j) \Delta^{(j-1)}_{\pm}} 
P 
e^{i(t_{j+1}-t_{j+2}) \Delta^{(j+1)}_{\pm}} \\
&=&e^{i(t_{j-1}-t_j) \Delta^{(j-1)}_{\pm}} 
e^{i(t_j-t_{j+1})(\Delta_{\pm, x_i})}
 B_{l, j} B_{i, j+1}\\
&& e^{i(t_j-t_{j+1})(\Delta^{(j)}_{\pm}-\Delta_{\pm, x_i})}
e^{i(t_{j+1}-t_{j+2}) \Delta^{(j+1)}_{\pm}} \\
 &=&e^{i(t_{j-1}- t_j)\Delta^{(j-1)}_{\pm}}
 e^{i(t_j-t_{j+1}) \Delta_{\pm, x_i}}
 B_{i, j+1} B_{l, j} \\
&\c&
e^{i(t_{j+1}-t_{j+2})(\Delta_{\pm, x_i}+\Delta_{\pm, x_{j+1}})}
e^{i(t_{j}-t_{j+2})(\Delta_{\pm, x_1} \cdots +\hat{\Delta}_{\pm, x_i}
+ \cdots 
+\Delta_{\pm, x_j})}
\eeaa
where a hat denotes a missing term.

Similarly, in view of the definition of 
$\tilde{\De}_{\pm}^{(j)}$, we can write,
\beaa
\tilde{\De}_{\pm}^{(j)}&=& \Delta_{\pm}^{(j)} -\Delta_{\pm, x_j} +\Delta_{\pm, x_{j+1}}\\
&=& \Delta_{\pm}^{(j-1)}+\Delta_{\pm, x_{j+1}}\\
&=&\Delta_{\pm}^{(j-1)}-\De_{{\pm},x_i}+\De_{{\pm},x_i}+\Delta_{\pm, x_{j+1}}
\eeaa
Hence,
\beaa
e^{-i(t_j-t_{j+1})\tilde{\Delta}_{\pm}^{(j)}}=e^{-i(t_j-t_{j+1})(\Delta_{\pm}^{(j-1)}-
\De{_\pm, x_i} ) } \,\c\, 
e^{-i(t_j-t_{j+1})( \Delta_{\pm, x_i}+\Delta_{\pm, x_{j+1}} )}
\eeaa
and consequently,
\beaa
\tilde P&=&
e^{-i(t_j-t_{j+1})(\Delta^{(j-1)}_{\pm}- \Delta_{\pm, x_i})}
 B_{i, j+1} B_{l, j}
 e^{-i(t_j-t_{j+1})(\Delta_{\pm, x_i}+\Delta_{\pm, x_{j+1}})}
\eeaa
Now,
\beaa
&&e^{i(t_{j-1}-t_{j+1})\Delta^{(j-1)}_{\pm}} \tilde P e^{i(t_{j}-t_{j+2}) \Delta^{(j+1)}_{\pm}}\\
&=&e^{i(t_{j-1}-t_{j+1})\Delta^{(j-1)}_{\pm}}e^{-i(t_j-t_{j+1})(\Delta^{(j-1)}_{\pm}- \Delta_{\pm, x_i})}
 B_{i, j+1} B_{l, j}\\
 &\c&
 e^{-i(t_j-t_{j+1})(\Delta_{\pm, x_i}+\Delta_{\pm, x_{j+1}})} e^{i(t_{j}-t_{j+2}) \Delta^{(j+1)}_{\pm}}\\
 &=&e^{i(t_{j-1}- t_j)\Delta^{(j-1)}_{\pm}}
 e^{i(t_j-t_{j+1}) \Delta_{\pm, x_i}}
 B_{i, j+1} B_{l, j} \\
&\c&
e^{i(t_{j+1}-t_{j+2})(\Delta_{\pm, x_i}+\Delta_{\pm, x_{j+1}})}
e^{i(t_{j}-t_{j+2})(\Delta_{\pm, x_1} \cdots +\hat{\Delta}_{\pm, x_i}
+ \cdots 
+\Delta_{\pm, x_j})}
\eeaa
and \eqref{comm} is proved.

Now the argument proceeds as in the example.
In the integral \eqref{firstint} use the symmetry \eqref{symm} to exchange
$x_j, x'_j$ with $x_{j+1}, x'_{j+1}$ 
in the arguments of $\gamma^{(n+1)}$ (only). Then use \eqref{comm}
in the integrand and also replace the $B's$  by their  corresponding integral kernels $\delta$.
Then we make the 
  change of  variables which  exchanges
$ t_j, x_j, x'_j$ with $t_{j+1}, x_{j+1}, x'_{j+1}$ in the whole integral.
To see the change in the domain of integration, say $\sigma(a)=j$ and
$\sigma(b)=j+1$, and say $b<a$. Then the domain $ t_1 \ge \cdots
\sigma(b) \ge \cdots \ge \sigma(a) \cdots $ changes to
$ t_1 \ge \cdots
\sigma(a) \ge \cdots \ge \sigma(b) \cdots $. In other words, $a =
\sigma^{-1}(j)$ and
$b = \sigma^{-1}(j+1)$ have been reversed.
This proves \eqref{I=I'}. 
\end{proof}

Next, we consider the subset $\{\mu_s\}\subset M$  of special, 
upper echelon, matrices
in which each highlighted element of a higher row is to
the left of each highlighted element
of a lower row. 
 Thus \eqref{equivmatrix2} is
in upper echelon form, and \eqref{equivmatrix1} is not. According to our
definition, the matrix

\begin{equation}
\begin{pmatrix} 
\bold{B_{1,2}}&\bold{B_{1,3}}&B_{1,4}&  B_{1, 5}\\
0&B_{2, 3}&\bold{B_{2, 4}} & B_{2, 5}\\
0&0 & B_{3, 4}& B_{3, 5}\\
0 & 0 & 0 &  \bold{B_{4, 5}}\\
\end{pmatrix}
\end{equation}

is also  in upper echelon form 

\begin{lemma} \label{moves}
For each element of $M$ there is a finite set
of acceptable moves which brings it to upper echelon form.
\end{lemma}

\begin{proof}
The strategy is to
start with the first row and do acceptable moves to bring all marked
entries in the first row in consecutive order, ${\bold B_{1, 2}}$
through ${\bold B_{1, k}}$. If there are any highlighted elements
on the second row, bring them to ${\bold B_{2, k+1}}$, ${\bold B_{2, l}}$.
 This will not affect
the marked entries of the first row.
If no entries are highlighted on the second row, leave it blank and move to
the third row.
 Continue to lower rows.
In the end, the matrix is reduced to an upper echelon form. \end{proof}

\begin{lemma} Let $C_n$ be the number of $n \times n$ special, upper
echelon matrices of the type discussed above.
Then $C_n \le 4^n$.
 
\end{lemma}
\begin{proof}
The proof consists of 2 steps. First dis-assemble the
original special matrix by ``lifting'' all marked entries
to the first row. This partitions the first row into subsets
$\{1, 2, \cdots k_1\}$, $\{k_1+1, \cdots, k_2\}$ etc. Let $P_n$ be the number
of such partitions. Look at the last subset of the partition. It can have $0$ elements, in which case
there is no last partition. This case contributes
precisely one partitions to the total number $P_n$.
If the last subset has $k$ elements then the remaining $n-k$,  can contribute exactly $P_{n-k}$
partitions. 
Thus  $P_n = 1 + P_1 + \cdots P_{n-1}$, and 
 therefore $P_n \le 2^n$ 
by induction.
In the second step we will re-assemble the upper echelon matrix by lowering 
$\{1, 2, \cdots k_1\}$ 
to the first used  row (we give up the
requirement that only the upper triangle is used, thus
 maybe counting more matrices)
$\{k_1+1, \cdots, k_2\}$ to the second used row etc.   Now suppose  that  we have exactly
$i$ subsets in a given  partition of the first 
row, which will be lowered in an
order-preserving way to the available n rows.
This can be done in exactly ${n}\choose{i}$ ways.
Thus $C_n \le P_n \sum_i \binom {n}{i} \le 4^n$. This is in agreement with
the combinatorial arguments of \cite{ESY1}. 
\end{proof}

\begin{theorem} \label{diagram}
Let $\mu_s$ be a special, upper echelon matrix, and
write $\mu \sim \mu_s$ if $\mu$ can be reduced to
$\mu_s$ in finitely many acceptable moves.
There
exists   $D$
a subset of $[0, t_1]^n$ such that
\bea
 \sum_{\mu \sim \mu_s}
\int_0^{t_1}\cdots \int_0^{t_n}J( \tb_{n+1};\,\mu) dt_2 \cdots dt_{n+1}=
\int_DJ( \tb_{n+1};\,\mu_s) dt_2 \cdots dt_{n+1}\nn\\
 \label{duham} 
\eea

\end{theorem}

\begin{proof}
Start with the integral,
\[
\II(\mu,\id)=
\int_0^{t_1}\cdots \int_0^{t_n}J( \tb_{n+1};\,\mu) dt_2 \cdots dt_{n+1}
\]
with its  corresponding matrix, \begin{equation} 
\begin{pmatrix} 
t_2 & t_3 & t_4 & \cdots & t_{n+1}\\
\bold{B_{\mu(2),2}}&B_{1,3}&\bold{B_{\mu(4),4}}& \cdots & B_{1, n+1}\\
0&\bold{B_{\mu(3), 3}}&B_{2, 4}& \cdots & B_{2, n+1}\\
0&0 & B_{3, 4}& \cdots & B_{3, n+1}\\
\cdots & \cdots & \cdots & \cdots & \cdots\\
0 & 0 & 0 & \cdots & B_{n, n+1}\\
\end{pmatrix}
\end{equation}
As in Lemma \eqref{moves}
perform finitely many acceptable moves on it, transforming the matrix determined by the pair 
$(\mu, \id)$ to the  special upper echelon form matrix corresponding   to a  pair $(\mu_s, \sigma)$,
\begin{equation} \label{matsigma}
\begin{pmatrix} 
t_{\sigma^{-1} (2)}&t_{\sigma^{-1}(3)}&t_{\sigma^{-1}(4)}& \cdots & t_{\sigma^{-1}(n+1)}\\
\bold{B_{1,2}}&\bold{B_{1,3}}&B_{1,4}& \cdots & B_{1, n+1}\\
0&B_{2, 3}&\bold{B_{2, 4}}& \cdots & B_{2, n+1}\\
0&0 & B_{3, 4}& \cdots & B_{3, n+1}\\
\cdots & \cdots & \cdots & \cdots & \cdots\\
\end{pmatrix}
\end{equation}
By Lemma \eqref{ints}, $\II(\mu, \id) = \II(\mu_s, \sigma)$.
Now observe that  if  $(\mu_1, id)$ and  $(\mu_2, id)$,
with $\mu_1\neq \mu_2$
lead to the same echelon form $\mu_s$ the corresponding permutations $\si_1$ and $\si_2$
must be different.
The lemma is thus  proved with $D$ the union of all
$\{t_1 \ge t_{\sigma(2)} \ge t_{\sigma(3)} \ge\cdots t_{\sigma(n_1)}\}$
for all permutations  $\sigma$ which occur
in a given class of equivalence of a given
 $\mu_s$.
 \end{proof}
\bigskip
\emph{Proof of Main  Theorem \eqref{maintheorem}}
\,\,\, 
 We start by fixing  $t_1$.
Express 
\bea
\gamma^{(1)}(t_1, \cdot)&=&\sum_{\mu }
\int_0^{t_1}\cdots \int_0^{t_n} J(\underline{t}_{n+1},\mu)
\eea
where, we recall,
\beaa
J(\underline{t}_{n+1},\mu)
&=&e^{i(t_1-t_2) \Delta^{(1)}_{\pm}}
B_{1,2}
e^{i(t_2-t_3) \Delta^{(2)}_{\pm}}
B_{\mu(3), 3}
\cdots \\
&&e^{i(t_n-t_{n+1}) \Delta^{(n)}_{\pm}}B_{\mu (n+1), n+1}
(\gamma^{(n+1)})(t_{n+1}, \cdot)
\eeaa
Using Theorem \eqref{diagram} we can write
$\gamma^{(1)}(t_1, \cdot)$ as a sum of at most
$4^n$ terms of the form
\begin{equation}
\int_DJ(\underline{t}_{n+1},\mu_s)
\end{equation}

Let $C^n=[0, t_1]\times [0, t_1] \times \cdots \times [0, t_1]$ 
(product of n terms).
Also, let $D_{t_2}= \{(t_3, \cdots,  t_{n+1})| 
(t_2, t_3, \cdots,   t_{n+1}) \in D\}$.
We have
\begin{align}
&\|R^{(1)} \gamma^{(1)}(t_1, \cdot)\|_{L^2(\RRR^3 \times \notag \RRR^3)}\\
&=\|R^{(1)}\int_D
e^{i(t_1-t_2) \Delta^{(1)}_{\pm}}
B_{1,2}
e^{i(t_2-t_3) \Delta^{(2)}_{\pm}}
B_{\mu_s(3), 3}
\cdots  \, dt_2 \cdots dt_{n+1}\|_{L^2(\RRR^3 \times \RRR^3)}
\notag\\
&= \| \int_0^{t_1} 
e^{i(t_1-t_2) \Delta^{(1)}_{\pm}}
\left(
\int_{D_{t_2}} 
R^{(1)}
B_{1,2} 
e^{i(t_2-t_3) \Delta^{(2)}_{\pm}}
B_{\mu_s(3), 3}
\cdots  \, dt_3 \cdots dt_{n+1} \right) dt_2\|_{L^2(\RRR^3 \times \RRR^3)}
\notag\\
&\le  \int_0^{t_1}  \|
e^{i(t_1-t_2) \Delta^{(1)}_{\pm}}
\int_{D_{t_2}} 
R^{(1)}
B_{1,2}\notag 
e^{i(t_2-t_3) \Delta^{(2)}_{\pm}}
B_{\mu_s(3), 3}
\cdots  \, dt_3 \cdots dt_{n+1} \|_{L^2(\RRR^3 \times \RRR^3)}
d t_2\\
&=  \int_0^{t_1}  \|
\int_{D_{t_2}} 
R^{(1)}
B_{1,2}\notag 
e^{i(t_2-t_3) \Delta^{(2)}_{\pm}}
B_{\mu_s(3), 3}
\cdots  \, dt_3 \cdots dt_{n+1} \|_{L^2(\RRR^3 \times \RRR^3)}
d t_2\\
&\le  \int_0^{t_1}  
\left(
\int_{D_{t_2}}\| 
R^{(1)}
B_{1,2}\notag 
e^{i(t_2-t_3) \Delta^{(2)}_{\pm}}
B_{\mu_s(3), 3}
\cdots 
 \|_{L^2(\RRR^3 \times \RRR^3)}
 \, dt_3 \cdots dt_{n+1} \right)
d t_2\\
&\le  
\int_{C^n}
\| R^{(1)}
B_{1,2}
e^{i(t_2-t_3) \Delta^{(2)}_{\pm}}
B_{\mu_s(3), 3}
\cdots 
\|_{L^2( \RRR^3 \times \RRR^3)}
 dt_2\, dt_3  \cdots dt_{n+1}
\notag\\
\end{align}
Applying  Cauchy-Schwarz in $t$ and Theorem \eqref{ourest}
  n-1 times , we estimate 
\begin{align}
&\int_{C^n}
\| R^{(1)}
B_{1,2}
e^{i(t_2-t_3) \Delta^{(2)}_{\pm}}
B_{\mu_s(3), 3}
\cdots 
\|_{L^2( \RRR^3 \times \RRR^3)}
 dt_2\, dt_3  \cdots dt_{n+1}
\notag\\
&\le t_1^{\frac{1}{2}} \int_{C^{n-1}}
\| R^{(1)}
B_{1,2}
e^{i(t_2-t_3) \Delta^{(2)}_{\pm}} \left(
B_{\mu_s(3), 3}
\cdots \right) 
\|_{L^2((t_2 \in  [0, t_1]) \times \RRR^3 \times \RRR^3)}
  dt_3  \cdots dt_{n+1}
\notag\\
&\le C t_1^{\frac{1}{2}} \int_{C^{n-1}}
\| R^{(2)}
B_{\mu_s(3), 3}
e^{i(t_3-t_4) \Delta^{(3)}_{\pm}}
B_{\mu_s(4), 4}
\cdots 
\|_{L^2(\RRR^6 \times \RRR^6)}
  dt_3  \cdots dt_{n+1}
\notag\\
& \cdots \notag\\
&\le (C t_1^{\frac{1}{2}} )^{n-1} 
\int_0^{t_1}\| R^{(n)}
B_{\mu_s(n+1), n+1}
\gamma^{(n+1)}(t_{n+1}, \cdot)\|_{L^2 (R^{3n} \times R^{3n})}
d t_{n+1}\notag\\
& \le  C  (C t_1^{\frac{1}{2}} )^{n-1} \notag
\end{align}
Consequently,
\bea
\|R^{(1)} \gamma^{(1)}(t_1, \cdot)\|_{L^2(\RRR^3 \times \RRR^3)}\le C  (C t_1^{\frac{1}{2}} )^{n-1}
\eea
If $C t_1 < 1$ and we let  $n \to \infty$ and infer that
$\|R^{(1)} \gamma^{(1)}(t_1, \cdot)\|_{L^2(\RRR^3 \times \RRR^3)}=0$.
 The proof for  all $\gamma^{(k)} =0$ is similar.
 Clearly we can continue the argument to show
 that all $\gamma^{(k)} $ vanish for all $t\ge 0$
 as desired.

\end{document}